\documentclass[12pt]{article}

\usepackage{amsmath}
\usepackage{amssymb}
\usepackage{latexsym}
\usepackage{graphicx}
\usepackage{epsfig}

\newcommand{\mysim}{\sim\!}

\newtheorem{theorem}{Theorem}[section]

\newtheorem{definition}[theorem]{Definition}

\newtheorem{lemma}[theorem]{Lemma}

\newtheorem{example}[theorem]{Example}

\newtheorem{corollary}[theorem]{Corollary}

\def\pushright#1{{\parfillskip=0pt\widowpenalty=10000
\displaywidowpenalty=10000\finalhyphendemerits=0\leavevmode\unskip
\nobreak\hfil\penalty50\hskip.2em\null\hfill{#1}\par}}

\def\qed{\xqed\global\SuppressEndOfProoftrue}
\newif\ifSuppressEndOfProof\SuppressEndOfProoffalse
\def\xqed{\pushright\markendofproof}
\def\markendofproof{\rule{1.3217ex}{2ex}}

\title{Minimum Model Semantics for Logic Programs\\ with Negation-as-Failure\footnote{
A preliminary version of this paper appears in the Proceedings of
the 8th European Conference on Logics in Artificial Intelligence (JELIA 2002),
Lecture Notes in Artificial Intelligence (LNAI) 2424, pages 456--467, Cosenza, Italy,
September 2002.}}

\author{Panos Rondogiannis\\
Department of Informatics \& Telecommunications\\
University of Athens\\
Panepistimiopolis, 157 84 Athens, Greece\\
e-mail: \textsf{\vspace{+0.3cm}prondo@di.uoa.gr}\\
\and 
William W. Wadge\\
Department of Computer Science\\
University of Victoria\\
PO Box 3055, STN CSC, Victoria, BC, Canada V8W 3P6\\
e-mail: \textsf{\vspace{+0.5cm}wwadge@csr.uvic.ca}
}

\date{}

\begin{document}
\maketitle

\vspace{-1.5cm}

\thispagestyle{empty}
\begin{abstract}
We give a purely model-theoretic characterization of
the semantics of logic programs with negation-as-failure allowed in clause bodies.
In our semantics the meaning of a program is, as in the classical case, 
the unique {\em minimum} model in a program-independent
ordering. We use an expanded truth domain that has an uncountable linearly
ordered set of truth values between {\em False} (the minimum element) and
{\em True} (the maximum), with a {\em Zero} element in the middle. The truth
values below {\em Zero} are ordered like the countable ordinals. The values
above {\em Zero} have exactly the reverse order. Negation is interpreted as
reflection about {\em Zero} followed by a step towards {\em Zero}; the only
truth value that remains unaffected by negation is {\em Zero}. We show that
every program has a unique minimum model $M_P$, and that this model can be
constructed with a $T_P$ iteration which proceeds through the countable ordinals.
Furthermore, we demonstrate that $M_P$ can also be obtained through a model
intersection construction which generalizes the well-known model intersection
theorem for classical logic programming. Finally, we show that by collapsing
the true and false values of the infinite-valued model $M_P$ to (the classical)
{\em True} and {\em False}, we obtain a three-valued model identical to the well-founded one.

\vspace{0.2cm}

\end{abstract}
\pagebreak

\pagestyle{plain}

\section{Introduction}
One of the paradoxes of logic programming is that such a small fragment of formal
logic serves as such a powerful programming language. This contrast has led to many attempts
to make the language more powerful by extending the fragment, but these
attempts generally back-fire. The extended languages can be implemented,
and are in a sense more powerful; but these extensions usually disrupt the
relationship between the meaning of programs {\em as programs} and the
meaning {\em as logic}. In these cases the implementation of the
program-as-program can no longer be considered as computing a distinguished
model of the program-as-logic. Even worse, the result of running the program may
not correspond to any model at all.

The problem is illustrated by the many attempts to extend logic programming
with negation (of atoms in the clause bodies). The generally accepted
computational interpretation of negated atoms is {\em negation-as-failure}.
Intuitively, a goal $\mysim A$ succeeds iff the subcomputation which attempts to
establish $A$ terminates and fails. Despite its simple computational formulation,
negation-as-failure proved to be extremely difficult to formalize from a 
semantic point of view (an overview of the existing semantic treatments
is given in the next section). Moreover, the existing approaches are not
purely model theoretic in the sense that the meaning of a given program can 
not be computed by solely considering its set of models. This is a sharp 
difference from classical logic programming (without negation), in which every program 
has a unique {\em minimum} Herbrand model (which is the intersection of all 
its Herbrand models).

This paper presents a purely model-theoretic semantics for 
negation-as-failure in logic programming. In our semantics the meaning of a program is,
as in the classical case, the unique minimum model in a program-independent
ordering. The main contributions of the paper can be summarized as follows: 
\begin{itemize}
\item We argue that a purely declarative semantics for logic programs with negation-as-failure 
      should be based on an infinite-valued logic. For this purpose we introduce an expanded
      truth domain that has an uncountable linearly ordered set of truth values between
      {\em False} (the minimum element) and {\em True} (the maximum), with a {\em Zero}
      element in the middle. The truth values below {\em Zero} are ordered like the countable
      ordinals while those above {\em Zero} have the reverse order. This new truth domain 
      allows us to define in a logical way the meaning of negation-as-failure and to 
      distinguish it in a very clear manner from classical negation.

\item We introduce the notions of {\em infinite-valued interpretation} and {\em infinite-valued
      model} for logic programs. Moreover, we define a partial ordering $\sqsubseteq_{\infty}$ on 
      infinite-valued interpretations which generalizes the subset ordering of classical 
      interpretations. We then demonstrate that every logic program that uses negation-as-failure,
      has a unique minimum (infinite-valued) model $M_P$ under $\sqsubseteq_{\infty}$. This model
      can be constructed by appropriately iterating a simple $T_P$ operator through the countable
      ordinals. From an algorithmic point of view, the construction of $M_P$ proceeds in an analogous 
      way as the {\em iterated least fixpoint} approach~\cite{P89-11}. There exist however crucial 
      differences. First and most important, the proposed approach aims at producing a unique minimum
      model of the program; this requirement leads to a more demanding logical setting than existing 
      approaches and the construction of $M_P$ is guided by the use of a family of relations on
      infinite-valued interpretations. Second, the definition of $T_P$ in the infinite-valued
      approach is a simple and natural extension of the corresponding well-known operator for 
      classical logic programming; in the existing approaches the operators used are complicated by 
      the need to keep track of the values produced at previous levels of the iteration. 
      Of course, the proposed approach is connected to the existing ones since, as we demonstrate, 
      if we collapse the true and false values of $M_P$ to (classical) {\em True} and 
      {\em False} we get the well-founded model.

\item We demonstrate that by considering infinite-valued models, we can derive 
      a model intersection theorem for logic programs with negation-as-failure.
      The model produced by the model intersection theorem coincides with the model 
      $M_P$ produced by $T_P$. To our knowledge, this is the first such result in 
      the area of negation (because model intersection does not hold if one restricts 
      attention to either two or three-valued semantical approaches). 
\end{itemize}
The rest of the paper is organized as follows: Section \ref{theproblem} discusses the
problem of negation and gives a brief outline of the most established semantic approaches.
Section \ref{intuitive} outlines the infinite-valued approach. Section \ref{infinite} 
introduces infinite-valued interpretations and models, and discusses certain orderings
on interpretations that will play a vital role in defining the infinite-valued semantics.
The $T_P$ operator on infinite-valued interpretations is defined in Section \ref{thetp}
and an important property of the operator, namely $\alpha$-monotonicity, is established. 
In Section \ref{themodel}, the construction of the model $M_P$ is presented. Section
\ref{properties} establishes various properties of $M_P$, the most important of which
is the fact that $M_P$ is the minimum model of $P$ under the ordering relation 
$\sqsubseteq_{\infty}$. Section \ref{intersection} introduces the model intersection
theorem and demonstrates that the model produced in this way is identical to $M_P$.
Finally, Section \ref{discussion} concludes the paper with discussion on certain aspects
of the infinite-valued approach.

\section{The Problem of Negation-as-Failure}\label{theproblem}
The semantics of negation-as-failure is possibly the most broadly studied
problem in the theory of logic programming. In this section we first discuss
the problem and then present the main solutions that have been proposed until now.

\subsection{The Problem}
Negation-as-failure is a notion that can be described operationally in a very simple
way, but whose denotational semantics has been extremely difficult to specify. This
appears to be a more general phenomenon in the theory of programming languages:
\begin{quote}
``It seems to be a general rule that programming language features and concepts
which are simple operationally tend to be complex denotationally, whereas those
which are simple denotationally tend to be complex operationally''~\cite{AW82-283}.
\end{quote}
The basic idea behind negation-as-failure has as follows: suppose that we are
given the goal $\leftarrow\,\, \mysim A$. Now, if $\leftarrow A$ succeeds, then
$\leftarrow\,\, \mysim A$ fails; if $\leftarrow A$ fails finitely, then $\leftarrow\,\, \mysim A$ 
succeeds. For example, given the program
\[
\begin{array}{lll}
 {\tt p} & \leftarrow & \\
 {\tt r} & \leftarrow & \mysim {\tt p}\\
 {\tt s} & \leftarrow & \mysim {\tt q}
\end{array}
\]
the query $\leftarrow {\tt r}$ fails because ${\tt p}$ succeeds, while $\leftarrow {\tt s}$
succeeds because ${\tt q}$ fails.

To illustrate the problems that result from the above interpretation of
negation, consider an even simpler program:
\[
\begin{array}{lll}
     {\tt works} & \leftarrow & \mysim {\tt tired}
\end{array}
\]
Under the negation-as-failure rule, the meaning of the above program is captured
by the model in which {\tt tired} is $False$ and {\tt works} is $True$.

Consider on the other hand the program:
\[
\begin{array}{lll}
     {\tt tired}& \leftarrow & \mysim {\tt works}
\end{array}
\]
In this case, the correct model under negation-as-failure is the one in which
{\tt works} is $False$ and {\tt tired} is $True$.

However, the above two programs have exactly the same classical models, namely:
\[
\begin{array}{lll}
  M_0 & = & \{({\tt tired},False),({\tt works},True)\}\\
  M_1 & = & \{({\tt tired},True),({\tt works},False)\}\\
  M_2 & = & \{({\tt tired},True),({\tt works},True)\}
\end{array}
\]

We therefore have a situation in which two programs have the same {\em model theory}
(set of models), but different computational meanings. Obviously, this implies that
the computational meaning does not have a purely model theoretic specification.
In other words, one can not determine the intended model of a logic program that uses 
negation-as-failure by just examining its set of models. This is a very sharp difference 
from logic programming without negation in which every program has a unique minimum model.

\subsection{The Existing Solutions}
The first attempt to give a semantics to negation-as-failure was the so-called
{\em program completion} approach introduced by Clark~\cite{C78}. In the completion
of a program the ``if'' rules are replaced by ``if and only if'' ones and also
an equality theory is added to the program (for a detailed presentation of the
technique, see \cite{lloyd}). The main problem is that the completion of a program
may in certain cases be inconsistent. To circumvent the problem, Fitting~\cite{F85-295}
considered 3-valued Herbrand models of the program completion. Later, Kunen~\cite{K87-289} 
identified a weaker version of Fitting's semantics which is recursively enumerable.
However, the last two approaches do not overcome all the objections that have been
raised regarding the completion (see for example the discussion in~\cite{PP90-321}
and in~\cite{vG93-185}).

Although the program completion approach proved useful in many application domains,
it has been superseded by other semantic approaches, usually termed under the name
{\em canonical model semantics}. The basic idea of the canonical model approach is to 
choose among the models of a program a particular one which is presumed to be the
model that the programmer had in mind. The canonical model is usually chosen among
many incomparable minimal models of the program. Since (as discussed in the last 
subsection) the selection of the canonical model can not be performed by just examining
the set of (classical) models of the program, the choice of the canonical model
is inevitably driven by the syntax of the program. In the following we discuss
the main semantic approaches that have resulted from this body of research.

A semantic construction that produces a single model is the so-called 
{\em stratified semantics}~\cite{ABW88}.  Informally speaking, a program is 
stratified if it does not contain cyclic dependencies of predicate names through 
negation. Every stratified logic program has a unique {\em perfect} model, which 
can be constructed in stages. As an example, consider again the program:
\[
\begin{array}{lll}
 {\tt p} & \leftarrow & \\
 {\tt r} & \leftarrow & \mysim {\tt p}\\
 {\tt s} & \leftarrow & \mysim {\tt q}
\end{array}
\]
The basic idea in the construction of the perfect model is to rank the predicate variables 
according to the maximum ``depth'' of negation used in their defining clauses. The 
variables of rank $0$ (like {\tt p} and {\tt q} above) are defined in terms of each
other without use of negation. The variables of rank $1$ (like {\tt r} and {\tt s}) are defined
in terms of each other and those of rank $0$, with negation applied only to variables of
rank $0$. Those of rank $2$ are defined with negations applied only to variables of rank
$1$ and $0$; and so on. The model can then be constructed in stages. The clauses for the 
rank $0$ variables form a standard logic program, and its minimum model is used to assign 
values for the rank $0$ variables. These are then treated as constants, so that the
clauses for the rank $1$ variables no longer have negations.
The minimum model is used to assign values to the rank $1$ variables, which
are in turn converted to constants; and so on. 

An extension of the notion of stratification is {\em local stratification}~\cite{Pri88}; 
intuitively, in a locally stratified program, predicates may depend negatively on themselves 
as long as no cycles are formed when the rules of the program are instantiated. Again, every 
locally stratified program has a unique perfect model~\cite{Pri88}. The construction of the
perfect model can be performed in an analogous way as in the stratified case (the basic
difference being that one can allow infinite countable ordinals as ranks).
It is worth noting that although stratification is obviously a syntactically
determinable condition, local stratification is generally undecidable~\cite{cholak}.
It should also be noted here that there exist some interesting cases of logic programming
languages where one can establish some intermediate notion between stratification
and local stratification which is powerful and decidable. For example, in temporal 
logic programming~\cite{mehmetsurvey,orgun92} many different temporal stratification
notions have been defined, and corresponding decision tests have been 
proposed~\cite{ZAO93,Lud98phd,prondotcs}.

The stratified and locally stratified semantics fail for
programs in which some variables are defined (directly or indirectly)
in terms of their own negations, because these variables are never ranked.
For such programs we need an extra intermediate neutral truth value for
certain of the negatively recursively defined variables. This approach yields
the ``well-founded'' construction and it can be shown~\cite{vGRS91-620}
that the result is indeed a model of the program. Many different constructive 
definitions of the well-founded model have been proposed; two of the most 
well-known ones are the {\em alternating fixpoint}~\cite{vG89-1,vG93-185} 
and the {\em iterated least fixpoint}~\cite{P89-11}. The well-founded
model approach is compatible with stratification (it is well-known that the 
well-founded model of a locally stratified program coincides with its unique 
perfect model~\cite{vGRS91-620}).

An approach that differs in philosophy from the previous ones is the so-called 
{\em stable model semantics}~\cite{GL88}.  While the ``canonical model'' approaches
assign to a given program a unique ``intended'' model, the stable model semantics
assigns to the program a (possibly empty) family of ``intended'' models.
For example, the program 
\[
\begin{array}{lll}
 {\tt p} & \leftarrow & \mysim {\tt p}
\end{array}
\]
does not have any stable models while the program
\[
\begin{array}{lll}
 {\tt p} & \leftarrow & \mysim {\tt q}\\
 {\tt q} & \leftarrow & \mysim {\tt p}
\end{array}
\]
has two stable models. The stable model semantics is defined through an elegant
{\em stability transformation}~\cite{GL88}. The relationships between the stable model
semantics and the previously mentioned canonical model approaches are quite close. It 
is well-known that every locally stratified program has a unique stable model which coincides 
with its unique perfect model~\cite{GL88}. Moreover, if a program has a two-valued well-founded 
model then this coincides with its unique stable model~\cite{vGRS91-620} (but the converse 
of this does not hold in general, see again~\cite{vGRS91-620}). Finally, as it is 
demonstrated in~\cite{P90-445}, the notion of stable model can be extended to a 
three-valued setting; then, the well-founded model can be characterized as the 
smallest (more precisely, the {\em F-least}, see~\cite{P90-445}) three-valued stable
model. The stable model approach has triggered the creation of a new promising 
programming paradigm, namely {\em answer-set programming}~\cite{MT99-375,GL02-3}.

It should be noted at this point that the infinite-valued approach proposed in this paper
contributes to the area of the ``canonical model'' approaches (and not in the area of
stable model semantics). In fact, as we argue in the next section, the infinite-valued
semantics is the purely model theoretic framework under which the existing canonical
model approaches fall.

The discussion in this section gives only a top-level presentation of the research that has been 
performed regarding the semantics of negation-as-failure. For a more in-depth treatment,
the interested reader should consult the many existing surveys for this area (such as
for example \cite{AB94,BG94-73,PP90-321,F02-25}).

\section{The Infinite-Valued Approach}\label{intuitive}

There is a general feeling (which we share) that when one seeks a unique model, then the 
well-founded semantics is the right approach to negation-as-failure. There still remains however 
a question about its legitimacy, mainly because the well-founded model is in fact one of the 
{\em minimal} models of the program and not a {\em minimum} one. In other words, there is 
nothing that distinguishes it {\em as a model}. 

Our goal is to remove the last doubts surrounding the well-founded model by providing 
a purely model theoretic semantics (the {\em infinite-valued semantics}) which is compatible 
with the well-founded model, but in which every program with negation has a unique minimum 
model. In our semantics whenever two programs have the same set of infinite-valued models then they 
have the same minimum model.

Informally, we extend the domain of truth values and use these extra values
to distinguish between ordinary negation and negation-as-failure, which we
see as being strictly weaker. Consider again the program:
\[
\begin{array}{lll}
 {\tt p} & \leftarrow & \\
 {\tt r} & \leftarrow & \mysim {\tt p}\\
 {\tt s} & \leftarrow & \mysim {\tt q}
\end{array}
\]
Under the negation-as-failure approach both {\tt p} and {\tt s} receive the value {\em True}. 
We would argue, however, that in some sense {\tt p} is ``truer'' than {\tt s}. Namely, {\tt p} 
is true because there is a rule which says so, whereas {\tt s} is true only because we 
are never obliged to make {\tt q} true. In a sense, {\tt s} is true only by default. Our truth 
domain adds a ``default'' truth value $T_1$ just below the ``real'' truth $T_0$, 
and (by symmetry) a weaker false value $F_1$ just above (``not as false as'') the real false 
$F_0$. We can then understand negation-as-failure as combining ordinary negation with a weakening. 
Thus $\mysim F_0 = T_1$ and $\mysim T_0 = F_1$. Since negations can effectively be iterated, our domain
requires a whole  sequence $\ldots,T_3, T_2, T_1$ of weaker and weaker truth values
below $T_0$ but above the neutral value $0$; and a mirror image sequence $F_1, F_2,F_3\ldots$
above $F_0$ and below $0$. In fact, to capture the well-founded model
in full generality, we need a $T_\alpha$ and a $F_\alpha$ for every countable ordinal $\alpha$.

We show that, over this extended domain, every logic program with negation
has a unique minimum model; and that in this model, if we collapse all the $T_\alpha$
and $F_\alpha$ to {\em True} and {\em False} respectively, we get the three-valued well-founded model.
For the example program above, the minimum model is 
$\{({\tt p},T_0),({\tt q},F_0),({\tt r},F_1),({\tt s},T_1)\}$.
This collapses to $\{({\tt p},True),({\tt q},False),({\tt r},False),({\tt s},True)\}$, which
is the well-founded model of the program. 

Consider now again the program ${\tt works} \leftarrow \mysim {\tt tired}$. The minimum model 
in this case is $\{({\tt tired},F_0),({\tt works},T_1)\}$. On the other hand, for the program
${\tt tired} \leftarrow \mysim {\tt works}$ the minimum model is $\{({\tt tired},T_1),({\tt works},F_0)\}$.
As it will become clearer in the next section, the minimum model of the 
first program is not a model of the second program, and vice-versa.
Therefore, the two programs do not have the same set of infinite-valued models
and the paradox identified in the previous section, disappears.
Alternatively, in the infinite-valued semantics the programs
${\tt works} \leftarrow \mysim {\tt tired}$ and ${\tt tired} \leftarrow \mysim {\tt works}$ 
are no longer logically equivalent.

The proof of our minimum-model result proceeds in a manner 
analogous to the classical proof in the negation-free case. 
The main complication is that we need extra auxiliary relations to
characterize the transitions between stages in the construction. This
complication is unavoidable and due to the fact that in our infinite 
truth domain negation-as-failure is {\em still} antimonotonic. 
The approximations do converge on the least model, but not monotonically
(or even anti-monotonically). Instead (speaking loosely) the values
of variables with standard denotations ($T_0$ and $F_0$) are computed 
first, then those ($T_1$ and $F_1$) one level weaker, then those
two levels weaker, and so on. We need a family of relations between
models to keep track of this intricate process (whose result, nevertheless, 
has a simple characterization).

\section{Infinite Valued Models}\label{infinite}
In this section we define infinite-valued interpretations and infinite-valued models 
of programs. In the following discussion we assume familiarity with the basic notions 
of logic programming~\cite{lloyd}. We consider the class of normal logic programs:
\begin{definition}
A {\em normal program clause} is a clause whose body is a conjunction of literals. A {\em normal
logic program} is a finite set of normal program clauses.
\end{definition}
We follow a common practice in the area of negation, which dictates that instead
of studying (finite) logic programs it is more convenient to study their (possibly
infinite) {\em ground instantiations}~\cite{F02-25}:
\begin{definition}
If $P$ is a normal logic program, its associated {\em ground instantiation} $P^*$ is constructed 
as follows: first, put in $P^*$ all ground instances of members of $P$; second, if
a clause $A \leftarrow$ with empty body occurs in $P^*$, replace it with $A \leftarrow \mbox{\tt true}$;
finally, if the ground atom $A$ is not the head of any member of $P^*$, add
$A \leftarrow \mbox{\tt false}$.
\end{definition}
The program $P^*$ is in essence a (generally infinite) propositional program.
In the rest of this paper, we will assume that all programs under consideration
(unless otherwise stated) are of this form.

The existing approaches to the semantics of negation are either 
two-valued or three-valued. The two-valued approaches are based on classical
logic that uses the truth values {\em False} and {\em True}. The three-valued
approaches are based on a three-valued logic that uses {\em False}, 0 and {\em True}.
The element 0 captures the notion of {\em undefined}. The truth values are
ordered as: \mbox{{\em False} $<$ 0 $<$ {\em True}} (see for example~\cite{P89-11}).

The basic idea behind the proposed approach is that in order to obtain a
minimum model semantics for logic programs with negation, it is
necessary to consider a much more refined multiple-valued logic which 
is based on an infinite set of truth values, ordered as follows:
$$F_0 < F_1 < \cdots < F_{\omega} < \cdots < F_{\alpha} < \cdots < 0 <
\cdots < T_\alpha < \cdots <T_\omega < \cdots <T_1 <T_0$$
Intuitively, $F_0$ and $T_0$ are the classical {\em False} and {\em True} values and
0 is the {\em undefined} value. The values below 0 are ordered like the countable ordinals. 
The values above 0 have exactly the reverse order. The intuition behind the new
values is that they express different levels of truthfulness 
and falsity. 
In the following we denote by $V$ the set consisting of the above truth values.
A notion that will prove useful in the sequel is that of the {\em order} of a given 
truth value:
\begin{definition}
The {\em order} of a truth value is defined as: $order(T_{\alpha}) = \alpha$,
$order(F_{\alpha}) = \alpha$ and $order(0) = +\infty$.
\end{definition}
The notion of ``Herbrand interpretation of a program'' can now be generalized:
\begin{definition}
An (infinite-valued) interpretation $I$ of a program $P$ is a function from the Herbrand Base $B_P$ of 
$P$ to $V$.
\end{definition}
In the rest of the paper, the term ``interpretation'' will mean an infinite-valued one
(unless otherwise stated).
As a special case of interpretation, we will use $\emptyset$ to denote the interpretation
that assigns the $F_0$ value to all atoms of a program.

In order to define the notion of {\em model} of a given program, we need to extend the 
notion of interpretation to apply to literals, to conjunctions of literals and
to the two constants {\tt true} and {\tt false} (for the purposes of 
this paper it is not actually needed to extend $I$ to more general formulas):
\begin{definition}\label{interpretation}
Let $I$ be an interpretation of a given program $P$. Then, $I$ can be extended 
as follows:
\begin{itemize}
\item For every negative atom $\mysim p$ appearing in $P$:
      \[ 
             I(\mysim p) = \left\{
                             \begin{array}{ll}
                             T_{\alpha + 1} & \mbox{if $I(p) = F_\alpha$}\\
                             F_{\alpha + 1} & \mbox{if $I(p) = T_\alpha$}\\
                             0              & \mbox{if $I(p) = 0$}
                             \end{array}
                      \right. 
      \]
\item For every conjunction of literals $l_1,\ldots,l_n$ appearing as
      the body of a clause in $P$: 
      \[
       I(l_1,\ldots,l_n) = min\{I(l_1),\ldots,I(l_n)\}
      \]
\end{itemize}
Moreover, $I(\mbox{\tt true}) = T_0$ and $I(\mbox{\tt false}) = F_0$.
\end{definition}

It is important to note that the above definition provides a purely logical characterization
of what negation-as-failure is; moreover, it clarifies the difference between classical negation
(which is simply reflection about $0$) and negation-as-failure (which is reflection about $0$
followed by a step towards $0$). The operational intuition behind the above definition is that
the more times a value is iterated through negation, the closer to zero it gets.

The notion of satisfiability of a clause can now be defined:
\begin{definition}
Let $P$ be a program and $I$ an interpretation of $P$. Then, $I$ {\em satisfies}
a clause $p \leftarrow l_1,\ldots,l_n$ of $P$ if $I(p) \geq I(l_1,\ldots,l_n)$. Moreover,
$I$ is a {\em model} of $P$ if $I$ satisfies all clauses of $P$.
\end{definition}
%
%
%
Given an interpretation of a program, we adopt specific notations
for the set of predicate symbols of the program that are assigned a 
specific truth value and for the subset of the interpretation that
corresponds to a particular order:
\begin{definition}
Let $P$ be a program, $I$ an interpretation of $P$ and $v\in V$. Then
$I \parallel v = \{p \in B_P \mid I(p) = v \}$. Moreover, if $\alpha$ is
a countable ordinal, then $I \sharp \alpha = \{(p,v) \in I \mid order(v) = \alpha \}$.
\end{definition}
The following relations on interpretations will prove useful
in the rest of the paper:
\begin{definition}
Let $I$ and $J$ be interpretations of a given program $P$ and $\alpha$ be a countable
ordinal. We write $I=_{\alpha} J$, if for all $\beta \leq a$, 
$I\parallel T_{\beta} = J\parallel T_{\beta}$
and $I\parallel F_{\beta} = J\parallel F_{\beta}$.
\end{definition}
\begin{example}
Let $I = \{({\tt p},T_0),({\tt q},T_1),({\tt r},T_2)\}$ and 
$J = \{({\tt p},T_0),({\tt q},T_1),({\tt r},F_2)\}$. Then,
$I =_{1} J$, but it is not the case that $I =_{2} J$.
\end{example}
\begin{definition}
Let $I$ and $J$ be interpretations of a given program $P$ and $\alpha$ be a countable ordinal. We write
$I\sqsubset_{\alpha} J$, if for all $\beta < a$, $I=_{\beta} J$ and
either  $I\parallel T_{\alpha} \subset J\parallel T_{\alpha}$ and $I\parallel F_{\alpha} \supseteq J\parallel F_{\alpha}$,
or $I\parallel T_{\alpha} \subseteq J\parallel T_{\alpha}$ and $I\parallel F_{\alpha} \supset J\parallel F_{\alpha}$.
We write $I \sqsubseteq_{\alpha} J$ if $I =_{\alpha} J$ or $I \sqsubset_{\alpha} J$.
\end{definition}
\begin{example}
Let $I = \{({\tt p},T_0),({\tt q},T_1),({\tt r},F_2)\}$ and $J = \{({\tt p},T_0),({\tt q},T_1),({\tt r},T_2)\}$.
Obviously, $I \sqsubset_{2} J$.
\end{example}
\begin{definition}
Let $I$ and $J$ be interpretations of a given program $P$. We write
$I\sqsubset_{\infty} J$, if there exists a countable ordinal $\alpha$ such that
$I \sqsubset_{\alpha} J$. We write $I\sqsubseteq_{\infty} J$ if either
$I = J$ or $I\sqsubset_{\infty} J$.
\end{definition}

It is easy to see that the relation $\sqsubseteq_{\infty}$ on the set of interpretations
of a given program, is a partial order (ie. it is reflexive, transitive and antisymmetric).
On the other hand, for every countable ordinal $\alpha$, the relation $\sqsubseteq_{\alpha}$ 
is a preorder (ie. reflexive and transitive). The following lemma gives a condition 
related to $\sqsubseteq_{\infty}$ which will be used in a later section:
\begin{lemma}\label{sqlemma}
Let $I$ and $J$ be two interpretations of a given program $P$. If for all $p$ in $P$
it is $I(p) \leq J(p)$, then $I\sqsubseteq_{\infty}J$.
\end{lemma}
\begin{proof}
If $I=J$ then obviously $I\sqsubseteq_{\infty}J$. Assume $I \neq J$ and let $\alpha$
be the least countable ordinal such that $I\sharp \alpha \neq  J\sharp \alpha$. Now,
for every $p$ in $P$ such that $J(p) = F_{\alpha}$, we have $I(p) \leq F_{\alpha}$.
However, since $I$ and $J$ agree on their values of order less than $\alpha$, we have
$I(p) = F_{\alpha}$. Therefore, $I \parallel F_{\alpha} \supseteq J \parallel F_{\alpha}$.
On the other hand, for every $p$ in $P$ such that $I(p) = T_{\alpha}$, we have 
$J(p) \geq T_{\alpha}$. Since $I$ and $J$ agree on their values of order less than $\alpha$,
we have $J(p) = T_{\alpha}$. Therefore, $I \parallel T_{\alpha} \subseteq J \parallel T_{\alpha}$.
Since $I\sharp \alpha \neq  J\sharp \alpha$, we get $I\sqsubset_{\alpha} J$ which implies
$I\sqsubseteq_{\infty}J$.
\end{proof}

The relation $\sqsubseteq_{\infty}$ will be used in the coming sections in order to
define the minimum model semantics for logic programs with negation-as-failure.
\begin{example}
Consider the program $P$:
\[
\begin{array}{lll}
 {\tt p} & \leftarrow & \mysim {\tt q}\\
 {\tt q} & \leftarrow & \mbox{\tt false}
\end{array}
\]
It can easily be seen that the interpretation $M_P = \{({\tt p},T_1),({\tt q},F_0)\}$ is the
least one (with respect to $\sqsubseteq_{\infty}$) among all infinite-valued
models of $P$. In other words, for every infinite-valued model $N$ of $P$,
it is $M_P \sqsubseteq_{\infty} N$.
\end{example}
We can now define a notion of monotonicity that will be the main tool in defining
the infinite-valued semantics: 
\begin{definition}
Let $P$ be a program and let $\alpha$ be a countable ordinal. A function $\Phi$ 
from the set of interpretations of $P$ to the set of interpretations of $P$ is called
{\em $\alpha$-monotonic} iff for all interpretations $I$ and $J$ of $P$, 
$I \sqsubseteq_{\alpha} J \Rightarrow \Phi(I) \sqsubseteq_{\alpha} \Phi(J)$.
\end{definition}
Based on the notions defined above, we can now define and examine the properties of
an immediate consequence operator for logic programs with negation-as-failure.

\section{The Immediate Consequence Operator}\label{thetp}

In this section we demonstrate that one can easily define a $T_P$ operator for
logic programs with negation, based on the notions developed in the last section. 
Moreover, we demonstrate that this operator is $\alpha$-monotonic
for all countable ordinals $\alpha$. The $\alpha$-monotonicity allows us to prove
that this new $T_P$ has a least fixpoint, for which however $\omega$
iterations are not sufficient. The procedure required for getting the least fixpoint
is more subtle than that for classical logic programs, and will be described shortly.
\begin{definition}
Let $P$ be a program and let $I$ be an interpretation of $P$.
The operator $T_P$ is defined as follows:\footnote{The notation $T_P(I)(p)$ is possibly
more familiar to people having some experience with functional programming: $T_P(I)(p)$
is the value assigned to $p$ by the interpretation $T_P(I)$.} 
    $$T_P(I)(p) = lub\{I(l_1,\ldots,l_n) \mid p\leftarrow l_1,\ldots,l_n \in P\}$$
$T_P$ is called the {\em immediate consequence operator} for $P$.
\end{definition}

The following lemma demonstrates that $T_P$ is well-defined:
\begin{lemma}
Every subset of the set $V$ of truth values has a least upper bound.
\end{lemma}
\begin{proof}
Let $V_F$ and $V_T$ be the subsets of $V$ that correspond to the false and true values respectively.
Let $S$ be a subset of $V$. Consider first the case in which $S\cap V_T$ is nonempty. 
Then, since $V_T$ is a reverse well-order, the subset $S \cap V_T$ must have a greatest element, 
which is clearly the least upper bound of $S$.

Now assume that $S\cap V_T$ is empty. Then, the intermediate truth value $0$ is an upper 
bound of $S$. If there are no other upper bounds in $V_F$, then $0$ is the least upper 
bound. But if the set of upper bounds of $S$ in $V_F$ is non empty, it must have a least 
element, because $V_F$ is well ordered; and this least element is clearly the least
upper bound of $S$ in the whole truth domain $V$.\qed
\end{proof}

\begin{example}
Consider the program:
\[
\begin{array}{lll}
 {\tt p} & \leftarrow & \mysim {\tt q}\\
 {\tt p} & \leftarrow &  \mysim {\tt p}\\
 {\tt q} & \leftarrow & \mbox{\tt false}
\end{array}
\]
and the interpretation $I = \{({\tt p},T_0),({\tt q},T_1)\}$. Then, $T_P(I) = \{({\tt p},F_2),({\tt q},F_0)\}$.
\end{example}
\begin{example}
For a more demanding example consider the following infinite program:
\[
\begin{array}{lllllllll}
{\tt p}_0       & \leftarrow & \mbox{\tt false}         &  &  &     & {\tt q}      & \leftarrow & {\tt p}_0 \\
{\tt p}_1       & \leftarrow & \mysim {\tt p}_0         &  &  &     & {\tt q}      & \leftarrow & {\tt p}_1 \\
{\tt p}_2       & \leftarrow & \mysim {\tt p}_1         &  &  &     & {\tt q}      & \leftarrow & {\tt p}_2 \\
{\tt p}_3       & \leftarrow & \mysim {\tt p}_2         &  &  &     & {\tt q}      & \leftarrow & {\tt p}_3 \\
                &   \ldots   &                          &  &  &     &              &   \ldots   &     
\end{array}
\]
Let $I = \{({\tt q},F_0),({\tt p}_0,F_0),({\tt p}_1,F_1),({\tt p}_2,F_2),\ldots\}$. Then, it can be easily seen
that $T_P(I) = \{({\tt q},F_{\omega}),({\tt p}_0,F_0),({\tt p}_1,T_1),({\tt p}_2,T_2),\ldots\}$.
\end{example}
One basic property of $T_P$ is that it is $\alpha$-monotonic, a property that is illustrated by
the following example:
\begin{example}
Consider the program:
\[
\begin{array}{lll}
 {\tt p} & \leftarrow & \mysim {\tt q}\\
 {\tt q} & \leftarrow & \mbox{\tt false}
\end{array}
\]
Let $I = \{({\tt q},F_0),({\tt p},T_2)\}$ and $J= \{({\tt q},F_1),({\tt p},T_0)\}$. 
Clearly, $I \sqsubseteq_0 J$. It can easily be seen that
$T_P(I) = \{({\tt q},F_0),({\tt p},T_1)\}$ and $T_P(J) = \{({\tt q},F_0),({\tt p},T_2)\}$,
and obviously $T_P(I) \sqsubseteq_0 T_P(J)$.
\end{example}
The following lemma establishes the $\alpha$-monotonicity of $T_P$. Notice that
a similar lemma also holds for the well-founded semantics (see for example~\cite{P89-11}). 
\begin{lemma}
The immediate consequence operator $T_P$ is $\alpha$-monotonic, for all countable ordinals $\alpha$.
\end{lemma}
\begin{proof}
The proof is by transfinite induction on $\alpha$. Assume the lemma holds for all
$\beta < \alpha$. We demonstrate that it also holds for $\alpha$. 

Let $I$, $J$ be two interpretations of $P$ such that $I\sqsubseteq_{\alpha } J$. 
We first establish that the values of order less that $\alpha$ remain intact by $T_P$.
Since $I\sqsubseteq_{\alpha } J$, for all $\beta < \alpha$ we have
$I\sqsubseteq_{\beta } J$ and $J\sqsubseteq_{\beta } I$. By the induction
hypothesis, we have that $T_P(I)\sqsubseteq_{\beta} T_P(J)$ and
$T_P(J)\sqsubseteq_{\beta} T_P(I)$, which implies that $T_P(I)=_{\beta} T_P(J)$, 
for all $\beta < \alpha$. It remains to show that $T_P(I) \parallel T_{\alpha} \subseteq T_P(J) \parallel T_{\alpha}$
and that $T_P(I) \parallel F_{\alpha} \supseteq T_P(J) \parallel F_{\alpha}$. We distinguish
these two cases.

We first demonstrate that $T_P(I) \parallel T_{\alpha} \subseteq T_P(J) \parallel T_{\alpha}$.
Assume that for some predicate $p$ in $P$ it is $T_P(I)(p) = T_{\alpha}$. 
We need to show that $T_P(J)(p) = T_{\alpha}$. Obviously, $T_P(J)(p) \leq T_{\alpha}$
since $T_P(I)=_{\beta} T_P(J)$, for all $\beta < \alpha$.
Consider now the fact that $T_P(I)(p) = T_{\alpha}$. This implies that
there exists a rule of the form $p \leftarrow q_1,\ldots,q_n,\mysim w_1,\ldots,\mysim w_m$
in $P$ whose body evaluates under $I$ to the value $T_{\alpha}$. This means that for all $q_i$, 
$1\leq i \leq n$, it is $I(q_i) \geq T_{\alpha}$ and for all $w_i$, $1\leq i \leq m$, 
it is $I(\mysim w_i) \geq T_{\alpha}$ (or equivalently, $I(w_i) < F_{\alpha}$). But then, 
since $I\sqsubseteq_{\alpha} J$, the evaluation of the body of the above rule under the interpretation 
$J$ also results to the value $T_{\alpha}$. This together with the fact that $T_P(J)(p) \leq T_{\alpha}$ 
allows us to conclude (using the definition of $T_P$) that $T_P(J)(p) = T_{\alpha}$.

It now remains to demonstrate that $T_P(I) \parallel F_{\alpha} \supseteq T_P(J) \parallel F_{\alpha}$.
Assume that for some predicate $p$ in $P$ it is $T_P(J)(p) = F_{\alpha}$.
We need to show that $T_P(I)(p) = F_{\alpha}$. Obviously, $T_P(I)(p) \geq F_{\alpha}$
since $T_P(I)=_{\beta} T_P(J)$, for all $\beta < \alpha$. Now, the fact that $T_P(J)(p) = F_{\alpha}$
implies that for every rule for $p$ in $P$, the body of the rule has a value under $J$ that is 
less than or equal to $F_{\alpha}$. Therefore, if $p \leftarrow q_1,\ldots,q_n,\mysim w_1,\ldots,\mysim w_m$
is one of these rules, then either there exists a $q_i$, $1\leq i \leq n$, such that
$J(q_i) \leq F_{\alpha}$, or there exists a $w_i$, $1\leq i \leq m$, such that $J(\mysim w_i) \leq F_{\alpha}$ 
(or equivalently $J(w_i) > T_{\alpha}$). But then, since $I\sqsubseteq_{\alpha} J$, the body of the
above rule evaluates under $I$ to a value less than or equal to $F_{\alpha}$. Therefore, 
$T_P(I)(p) \leq F_{\alpha}$. This together with the fact that $T_P(J)(p) \geq F_{\alpha}$ 
imply that $T_P(J)(p) = F_{\alpha}$.

\qed
\end{proof}

It is natural to wonder whether $T_P$ is monotonic with respect to the 
relation $\sqsubseteq_{\infty}$. This is not the case, as the following example
illustrates:
\begin{example}
Consider the program:
\[
\begin{array}{lll}
 {\tt p} & \leftarrow & \mysim {\tt q}\\
 {\tt s} & \leftarrow & {\tt p}\\
 {\tt t} & \leftarrow & \mysim {\tt s}\\
 {\tt t} & \leftarrow & {\tt u}\\
 {\tt u} & \leftarrow & {\tt t}\\
 {\tt q} & \leftarrow & \mbox{\tt false}
\end{array}
\]
Consider the following interpretations: $I = \{({\tt p},T_1),({\tt q},F_0),({\tt s},F_0),({\tt t},T_1),({\tt u},F_0)\}$ and 
$J = \{({\tt p},T_1),({\tt q},F_0),({\tt s},F_1),({\tt t},F_1),({\tt u},F_1)\}$. Obviously, it is $I\sqsubseteq_{\infty} J$ because
$I\sqsubset_{0} J$. However, we have $T_P(I) = \{({\tt p},T_1),({\tt q},F_0),({\tt s},T_1),({\tt t},T_1),({\tt u},T_1)\}$ and 
also $T_P(J) = \{({\tt p},T_1),({\tt q},F_0),({\tt s},T_1),({\tt t},T_2),({\tt u},F_1)\}$.  
Clearly, $T_P(I)\not\sqsubseteq_{\infty}T_P(J)$. 
\end{example}

The fact that $T_P$ is not monotonic under $\sqsubseteq_{\infty}$ appears to suggest that if we 
want to find the least (with respect to $\sqsubseteq_{\infty}$) fixpoint of $T_P$, we should 
not rely on approximations based on the relation $\sqsubseteq_{\infty}$. The way that this minimum 
fixpoint can be constructed, is described in the following section.

\section{Construction of the Minimum Model $M_P$}\label{themodel}
In this section we demonstrate how the minimum model $M_P$ of a given program $P$ can be
constructed. The construction can informally be described as follows. As a first 
approximation to $M_P$, we start with the interpretation that assigns to every atom of 
$P$ the value $F_0$ (as already mentioned, this interpretation is denoted by $\emptyset$). 
We start iterating the $T_P$ on $\emptyset$ until both the set of atoms that have a $F_0$
value and the set of atoms having a $T_0$ value, stabilize. We keep all these atoms whose 
values have stabilized and reset the values of all remaining atoms to the next false
value (namely $F_1$). The procedure is repeated until the $F_1$ and $T_1$ values stabilize,
and we reset the remaining atoms to a value equal to $F_2$, and so on. Since the
Herbrand Base of $P$ is countable, there exists a countable ordinal $\delta$ for which this
process will not produce any new atoms having $F_\delta$ or $T_\delta$ values. At this
point we stop the iterations and reset all remaining atoms to the value 0. The above
process is illustrated by the following example:
\begin{example}\label{mainexample}
Consider the program:
\[
\begin{array}{lll}
 {\tt p} & \leftarrow & \mysim {\tt q}\\
 {\tt q} & \leftarrow & \mysim {\tt r}\\
 {\tt s} & \leftarrow & {\tt p}\\
 {\tt s} & \leftarrow & \mysim {\tt s} \\
 {\tt r} & \leftarrow & \mbox{\tt false}
\end{array}
\]
We start from the interpretation $I = \{({\tt p},F_0),({\tt q},F_0),({\tt r},F_0),({\tt s},F_0)\}$. Iterating
the immediate consequence operator twice, we get in turn the following two interpretations:
\[
\begin{array}{l}
\{({\tt p},T_1),({\tt q},T_1),({\tt r},F_0),({\tt s},T_1)\}\\
\{({\tt p},F_2),({\tt q},T_1),({\tt r},F_0),({\tt s},T_1)\}
\end{array}
\]
Notice that the set of atoms having an $F_0$ value as well as the set of atoms having a $T_0$ value,
have stabilized (there is only one atom having an $F_0$ value and none having a $T_0$ one). Therefore, 
we reset the values of all other atoms to $F_1$ and repeat the process until the $F_1$ and $T_1$ values 
converge:
\[
\begin{array}{l}
\{({\tt p},F_1),({\tt q},F_1),({\tt r},F_0),({\tt s},F_1)\}\\
\{({\tt p},T_2),({\tt q},T_1),({\tt r},F_0),({\tt s},T_2)\}\\
\{({\tt p},F_2),({\tt q},T_1),({\tt r},F_0),({\tt s},T_2)\}
\end{array}
\]
Now, the order $1$ values have converged, so we reset all remaining values to $F_2$
and continue the iterations:
\[
\begin{array}{l}
\{({\tt p},F_2),({\tt q},T_1),({\tt r},F_0),({\tt s},F_2)\}\\
\{({\tt p},F_2),({\tt q},T_1),({\tt r},F_0),({\tt s},T_3)\}\\
\{({\tt p},F_2),({\tt q},T_1),({\tt r},F_0),({\tt s},F_4)\}
\end{array}
\]
The order $2$ values have converged, and we reset the value of {\tt s} to $F_3$:
\[
\begin{array}{l}
\{({\tt p},F_2),({\tt q},T_1),({\tt r},F_0),({\tt s},F_3)\}\\
\{({\tt p},F_2),({\tt q},T_1),({\tt r},F_0),({\tt s},T_4)\}
\end{array}
\]
The fact that we do not get any order $3$ value implies that we have reached the
end of the iterations. The final model results by setting the value of {\tt s} to $0$:
$$M_P = \{({\tt p},F_2),({\tt q},T_1),({\tt r},F_0),({\tt s},0)\}$$ 
As it will be demonstrated, this is the minimum model of the program under $\sqsubseteq_{\infty}$.
\end{example}
The above notions are formalized by the definitions that follow.
\begin{definition}
Let $P$ be a program, let $I$ be an interpretation of $P$ and $\alpha$ a
countable ordinal. Moreover, assume that $I \sqsubseteq_{\alpha} T_P(I) 
\sqsubseteq_{\alpha} T_P^2(I) \sqsubseteq_{\alpha} \cdots
\sqsubseteq_{\alpha} T_P^n(I) \sqsubseteq_{\alpha} \cdots$, $n < \omega$.
Then, the sequence $\{T_P^n(I)\}_{n < \omega}$ is called an $\alpha$-chain.
\end{definition}
\begin{definition}
Let $P$ be a program, let $I$ be an interpretation of $P$ and assume that
$\{T_P^n(I)\}_{n < \omega}$ is an $\alpha$-chain. Then, we define the
interpretation $T_{P,\alpha}^{\omega}(I)$ as follows:
      \[
             T_{P,\alpha}^{\omega}(I)(p) = \left\{
                             \begin{array}{ll}
                             I(p)           & \mbox{if $order(I(p)) < \alpha$}\\
                             T_{\alpha}     & \mbox{if $p\in \bigcup_{n < \omega}(T_P^n(I) \parallel T_{\alpha})$}\\
                             F_{\alpha}     & \mbox{if $p\in \bigcap_{n < \omega}(T_P^n(I) \parallel F_{\alpha})$}\\
                             F_{\alpha+1}   & \mbox{otherwise}
                             \end{array}
                      \right.
      \]
\end{definition}
The proof of the following lemma follows directly from the above definition:
\begin{lemma}\label{lublemma}
Let $P$ be a program, $I$ an interpretation of $P$ and $\alpha$ a countable
ordinal. Assume that $\{T_P^n(I)\}_{n < \omega}$ is an $\alpha$-chain.
Then, for all $n< \omega$, $T_P^{n}(I) \sqsubseteq_{\alpha}T_{P,\alpha}^{\omega}(I)$.
Moreover, for all interpretations $J$ such that for all $n<\omega$, 
$T_P^{n}(I) \sqsubseteq_{\alpha} J$, it is 
$T_{P,\alpha}^{\omega}(I) \sqsubseteq_{\alpha} J$.
\end{lemma}
The following definition and lemma will be used later on to suggest that the interpretations that 
result during the construction of the minimum model, do not assign to variables values of the form
$T_{\alpha}$ where $\alpha$ is a limit ordinal.
\begin{definition}
An interpretation $I$ of a given program $P$ is called {\em reasonable} if for all
$(p,T_{\alpha}) \in I$, $\alpha$ is not a limit ordinal.
\end{definition}
\begin{lemma}\label{nolimit}
Let $P$ be a program and $I$ a reasonable interpretation of $P$. Then, for all
$n < \omega$, $T_P^n(I)$ is a reasonable interpretation of $P$. Moreover, if
$\{T_P^n(I)\}_{n < \omega}$ is an $\alpha$-chain, then $T_{P,\alpha}^{\omega}(I)$
is a reasonable interpretation of $P$.
\end{lemma}
\begin{proof}
The proof of the first part of the theorem is by induction on $n$. For $n=0$
the result is immediate. Assume that $T_P^k(I)$ is reasonable,
and consider the case of $T_P^{k+1}(I)$. Now, if $(p,T_{\alpha})$ belongs to 
$T_P^{k+1}(I)$, where $\alpha$ is a limit ordinal, then there must exist a clause
$p \leftarrow B$ in $P$ such that $T_P^{k}(I)(B) = T_{\alpha}$. But this implies
that there exists a literal $l$ in $B$ such that $T_P^{k}(I)(l) = T_{\alpha}$.
If $l$ is a positive literal, then this is impossible due to the induction
hypothesis. If $l$ is a negative literal, this is impossible from the interpretation
of $\sim$ in Definition \ref{interpretation}.

The proof of the second part of the theorem is immediate: if $(p,T_{\alpha}) \in T_{P,\alpha}^{\omega}(I)$
then (by the definition of $T_{P,\alpha}^{\omega}$) there exists $k < \omega$ such that  
$(p,T_{\alpha}) \in T_P^{k}(I)$. But this is impossible from the first part
of the theorem. \qed
\end{proof}
We now define a sequence of interpretations of a given program $P$ (which can be thought
of as better and better approximations to the minimum model of $P$):
\begin{definition}\label{approx}
Let $P$ be a program and let:
\[
\begin{array}{ccll}
   M_0 & = & T^{\omega}_{P,0}(\emptyset) & \\
   M_{\alpha} & = & T^{\omega}_{P,\alpha}(M_{\alpha - 1})& \mbox{for successor ordinal $\alpha$}\\
   M_{\alpha} & = & T^{\omega}_{P,\alpha}(\bigsqcup_{\beta < \alpha} M_{\beta})& \mbox{for limit ordinal $\alpha$}
\end{array}
\]
where:
\[
       (\bigsqcup_{\beta < \alpha} M_{\beta})(p) = \left\{
                    \begin{array}{ll}
                      (\bigcup_{\beta < \alpha}(M_{\beta}\sharp \beta))(p)& \mbox{if this is defined}\\
                      F_{\alpha}& \mbox{otherwise}
                    \end{array}
                  \right.
\]
The $M_0,M_1,\ldots,M_{\alpha},\ldots$ are called the {\em approximations} to the minimum 
model of $P$.
\end{definition}
From the above definition it is not immediately obvious that the approximations are
well-defined. First, the definition of $T^{\omega}_{P,\alpha}$ presupposes the existence
of an $\alpha$-chain (for example, in the definition of $M_0$ one has to demonstrate that
$\{T_P^n(\emptyset)\}_{n < \omega}$ is a $0$-chain). Second, in the definition of 
$\bigsqcup_{\beta < \alpha} M_{\beta}$ above, we implicitly assume that 
$\bigcup_{\beta < \alpha}(M_{\beta}\sharp \beta)$ is a function. But in order
to establish this, we have to demonstrate that the domains of the relations
$M_{\beta}\sharp \beta$, $\beta < \alpha$, are disjoint (ie. that no predicate 
name participates simultaneously to more than one $M_{\beta}\sharp \beta$).
The following lemma clarifies the above situation. Notice that the lemma consists
of two parts, which are proven simultaneously by transfinite induction. This is
because the induction hypothesis of the second part is used in the induction step 
of the first part.
\begin{lemma}\label{fixlemma}
For all countable ordinals $\alpha$:
\begin{enumerate}
\item $M_{\alpha}$ is well-defined, and
\item $T_P(M_{\alpha}) =_{\alpha} M_{\alpha}$. 
\end{enumerate}
\end{lemma}
\begin{proof} 
The proof is by transfinite induction on $\alpha$. We distinguish three cases:

{\em Case 1:} $\alpha = 0$. In order to establish that the sequence 
$\{T_P^{n}(\emptyset)\}_{n< \omega}$ is a $0$-chain, we use induction
on $n$. For the basis case observe that $\emptyset \sqsubseteq_0 T_P(\emptyset)$. 
Moreover, if we assume that $T_P^n(\emptyset) \sqsubseteq_0 T_P^{n+1}(\emptyset)$,
using the $0$-monotonicity of $T_P$ we get that 
$T_P^{n+1}(\emptyset) \sqsubseteq_0 T_P^{n+2}(\emptyset)$.
Therefore, for all $n<\omega$, $T_P^n(\emptyset) \sqsubseteq_0 T_P^{n+1}(\emptyset)$.
It remains to establish that $T_P(M_{0}) =_{0} M_{0}$.

From Lemma \ref{lublemma}, $T_P^n(\emptyset) \sqsubseteq_0 M_0$, for all $n$. By the $0$-monotonicity 
of $T_P$, we have that for all $n<\omega$, $T_P^{n+1}(\emptyset) \sqsubseteq_0 T_P(M_0)$;
moreover, obviously $\emptyset \sqsubseteq_0 T_P(M_0)$. Therefore, for all $n < \omega$,
$T_P^{n}(\emptyset) \sqsubseteq_0 T_P(M_0)$. But then, from the second part of Lemma \ref{lublemma}, 
$M_0 \sqsubseteq_0 T_P(M_0)$. It remains to show that $T_P(M_0) \sqsubseteq_0 M_0$. Let $p$ be a 
predicate in $P$ such that $M_0(p) = F_0$. Then, for all $n$, $T_P^n(\emptyset)(p) = F_0$. This 
means that for every clause of the form $p \leftarrow B$ in $P$ and for all $n<\omega$, 
$T_P^{n}(\emptyset)(B) = F_0$. This implies that there exists a literal $l$ in $B$ such 
that for all $n < \omega$, $T_P^{n}(\emptyset)(l) = F_0$ (this is easily implied by the 
fact that $\{T_P^{n}(\emptyset)\}_{n< \omega}$ is a $0$-chain). Therefore, $M_0(l) = F_0$ 
and consequently $M_0(B) = F_0$, which shows that $T_P(M_0)(p) = F_0$. Consider on the other 
hand a predicate $p$ in $P$ such that $T_P(M_0) = T_0$. Then, there exists
a clause $p \leftarrow B$ in $P$ such that $M_0(B) = T_0$. This implies that for 
all literals $l$ in $B$, $M_0(l) = T_0$. But then there exists a $k$ such that for 
all $l$ in $B$ and all $n\geq k$, $T_P^n(\emptyset)(l) = T_0$ (this again is implied 
by the fact that $\{T_P^{n}(\emptyset)\}_{n< \omega}$ is a $0$-chain). This implies
that for all $n\geq k$, $T_P^n(\emptyset)(B) = T_0$ which means that
for all $n \geq k$, $T_P^{n+1}(\emptyset)(p) = T_0$. Consequently, $M_0(p) = T_0$.

\vspace{0.2cm}

{\em Case 2:} $\alpha$ is a limit ordinal. Then, 
$M_{\alpha} = T^{\omega}_{P,\alpha}(\bigsqcup_{\beta < \alpha} M_{\beta})$.
Based on the induction hypothesis one can easily verify that the domains of 
the relations $M_{\beta}\sharp \beta$, $\beta < \alpha$, are disjoint
and therefore the quantity $\bigsqcup_{\beta < \alpha} M_{\beta}$ is well-defined
(intuitively, the values of order less than or equal to $\beta$ in $M_{\beta}$
have stabilized and will not change by subsequent iterations of $T_P$). 
Moreover, it is easy to see that the sequence 
$\{T_P^{n}(\bigsqcup_{\beta < \alpha} M_{\beta})\}_{n< \omega}$ 
is an $\alpha$-chain (the proof is by induction 
on $n$ and uses the $\alpha$-monotonicity of $T_P$). 

It remains to establish that $T_P(M_{\alpha}) =_{\alpha} M_{\alpha}$.
We first show that $M_{\alpha} \sqsubseteq_{\alpha} T_P(M_{\alpha})$. 
Since $\{T_P^{n}(\bigsqcup_{\beta < \alpha} M_{\beta})\}_{n< \omega}$ is an
$\alpha$-chain, from Lemma \ref{lublemma}, 
$T_P^n(\bigsqcup_{\beta < \alpha} M_{\beta}) \sqsubseteq_{\alpha} M_{\alpha}$,
for all $n<\omega$. By the $\alpha$-monotonicity of $T_P$ we have that for all $n<\omega$,
$T_P^{n+1}(\bigsqcup_{\beta < \alpha} M_{\beta}) \sqsubseteq_{\alpha} T_P(M_{\alpha})$;
moreover, it is $\bigsqcup_{\beta < \alpha} M_{\beta} \sqsubseteq_{\alpha} T_P(M_{\alpha})$
(because $\bigsqcup_{\beta < \alpha} M_{\beta} \sqsubseteq_{\alpha} 
T_P(\bigsqcup_{\beta < \alpha} M_{\beta})$ and
$T_P(\bigsqcup_{\beta < \alpha} M_{\beta}) \sqsubseteq_{\alpha} T_P(M_{\alpha})$).
Therefore, for all $n<\omega$, $T_P^{n}(\bigsqcup_{\beta < \alpha} M_{\beta}) \sqsubseteq_{\alpha} T_P(M_{\alpha})$.
But then, by Lemma \ref{lublemma}, $M_{\alpha} \sqsubseteq_{\alpha} T_P(M_{\alpha})$.
Notice that this (due to the definition of $\sqsubseteq_{\alpha}$) immediately implies
that for all $\beta < \alpha$, $M_{\alpha} =_{\beta} T_{P}(M_{\alpha})$.

It remains to show that $T_P(M_{\alpha}) \sqsubseteq_{\alpha} M_{\alpha}$. 
It suffices to show that $T_P(M_{\alpha}) \parallel T_{\alpha} \subseteq M_{\alpha} \parallel T_{\alpha}$
and $T_P(M_{\alpha}) \parallel F_{\alpha} \supseteq M_{\alpha} \parallel F_{\alpha}$.
The former statement is immediate since (by Lemma \ref{nolimit}) values of the form
$T_{\alpha}$, where $\alpha$ is a limit ordinal, do not arise. Consider now the
latter statement and let $p$ be a predicate in $P$ such that $M_{\alpha}(p) = F_{\alpha}$. 
Then, by the definition of $T_{P,\alpha}^{\omega}$, we get that for all $n \geq 0$,
$T_P^n(\bigsqcup_{\beta < \alpha} M_{\beta})(p) = F_{\alpha}$. Assume that $T_P(M_{\alpha})(p) \neq F_{\alpha}$.
Then, since $M_{\alpha} =_{\beta} T_{P}(M_{\alpha})$ for all $\beta < \alpha$, it has to be
$T_P(M_{\alpha})(p) > F_{\alpha}$. But then this means that there exists a clause $p \leftarrow B$
in $P$ such that $M_{\alpha}(B) > F_{\alpha}$. This implies that for every literal
$l$ in $B$, it is $M_{\alpha}(l) > F_{\alpha}$. But then, by a case analysis on the possible
values that $M_{\alpha}(l)$ may have, one can show that there exists a $k$ such that for all
$l$ in $B$ and for all $n\geq k$, $T_P^n(\bigsqcup_{\beta < \alpha} M_{\beta})(l) > F_{\alpha}$.
In other words, for this particular clause there exists a $k$ such that for all $n \geq k$,
$T_P^n(\bigsqcup_{\beta < \alpha} M_{\beta})(B) > F_{\alpha}$. But this implies that 
for all $n \geq k$, $T_P^{n+1}(\bigsqcup_{\beta < \alpha} M_{\beta})(p) > F_{\alpha}$
(contradiction). Therefore, $T_{P}(M_{\alpha})(p) = F_{\alpha}$.

\vspace{0.2cm}

{\em Case 3:} $\alpha$ is a successor ordinal. Then, $M_{\alpha} = T_{P,\alpha}^{\omega}(M_{\alpha - 1})$.
As before, it is straightforward to establish that $\{T_P^{n}(M_{\alpha - 1})\}_{n< \omega}$ is an
$\alpha$-chain. Moreover, demonstrating that $M_{\alpha} \sqsubseteq_{\alpha} T_{P}(M_{\alpha})$ 
is performed in an entirely analogous way as in Case 2. Notice that this (due to the definition of $\sqsubseteq_{\alpha}$) 
immediately implies that for all $\beta < \alpha$, $M_{\alpha} =_{\beta} T_{P}(M_{\alpha})$.

It remains to show that $T_{P}(M_{\alpha}) \sqsubseteq_{\alpha}  M_{\alpha}$. For this, it suffices to establish 
that $T_P(M_{\alpha}) \parallel T_{\alpha} \subseteq M_{\alpha} \parallel T_{\alpha}$
and $T_P(M_{\alpha}) \parallel F_{\alpha} \supseteq M_{\alpha} \parallel F_{\alpha}$. Consider the former
statement and let $T_{P}(M_{\alpha})(p) = T_{\alpha}$, for some predicate $p$ in $P$. 
Then, since $M_{\alpha} =_{\beta} T_{P}(M_{\alpha})$ for all $\beta < \alpha$, it has to be
$M_{\alpha}(p) \leq T_{\alpha}$. Moreover, since $T_{P}(M_{\alpha})(p) = T_{\alpha}$, there exists
a clause $p \leftarrow B$ in $P$ such that $M_{\alpha}(B) = T_{\alpha}$. This implies that for every 
literal $l$ in $B$, $M_{\alpha}(l) \geq T_{\alpha}$. By a case analysis on the possible values that 
$M_{\alpha}(l)$ may have, one can show that there exists a $k$ such that for all $n \geq k$,
$T_P^n(M_{\alpha - 1})(l) = M_{\alpha}(l)$. This implies that for all $n \geq k$,
$T_P^{n}(M_{\alpha - 1})(B) = M_{\alpha}(B) =  T_{\alpha}$. This implies that
for all $n \geq k$, $T_P^{n+1}(M_{\alpha - 1})(p)\geq T_{\alpha}$ and therefore
$M_{\alpha}(p) \geq T_{\alpha}$. Now, since $M_{\alpha}(p) \leq T_{\alpha}$,
we conclude that $M_{\alpha}(p) = T_{\alpha}$.

The proof for the latter part of the statement is similar to the corresponding proof for Case 2.

\qed
\end{proof}

The following two lemmas are now needed in order to define the minimum model of a given
program:
\begin{lemma}
Let $P$ be a program. Then, there exists a countable ordinal $\delta$ such that:
\begin{enumerate}
\item $M_{\delta} \parallel T_{\delta} = \emptyset$ and $M_{\delta} \parallel F_{\delta} = \emptyset$
\item for all $\beta < \delta$, $M_{\beta} \parallel T_{\beta} \neq \emptyset$ or $M_{\beta} \parallel F_{\beta} \neq \emptyset$
\end{enumerate}
This ordinal $\delta$ is called the {\em depth} of $P$\footnote{The term ``depth''
was first used by T. Przymusinski in~\cite{P89-11}.}.
\end{lemma}
\begin{proof}
The basic idea behind the proof is that since $B_P$ is countable and the set of
countable ordinals is uncountable, there can not exist an onto function from the 
former set to the latter. More specifically, consider the set $S$ of pairs of truth
values of the form $(T_{\alpha},F_{\alpha})$, for all countable ordinals $\alpha$.
Consider the function $F$ that maps each predicate symbol $p \in B_P$ to 
$(T_{\alpha},F_{\alpha})$ if and only if $p \in M_{\alpha} \parallel F_{\alpha} \cup
M_{\alpha} \parallel T_{\alpha}$. Assume now that there does not exist a $\delta$
having the properties specified by the theorem. This would imply that every member
of the range of $F$ would be the map of at least one element from $B_P$. But this 
is impossible since $B_P$ is countable while the set $S$ is uncountable. To complete
the proof, take as $\delta$ the smallest countable ordinal $\alpha$ such that 
$M_{\alpha} \parallel T_{\alpha} = \emptyset$ and $M_{\alpha} \parallel F_{\alpha} = \emptyset$.
\qed
\end{proof}
The following property of $\delta$ reassures us that the approximations beyond $M_{\delta}$
do not introduce any new truth values:
\begin{lemma}\label{emptylemma}
Let $P$ be a program. Then, for all countable ordinals $\gamma \geq \delta$,
$M_{\gamma} \parallel T_{\gamma} = \emptyset$ and $M_{\gamma} \parallel F_{\gamma} = \emptyset$.
\end{lemma}
\begin{proof}(Outline)
The proof is by transfinite induction on $\gamma$. The basic idea is that 
if either $M_{\gamma} \parallel T_{\gamma}$  (respectively $M_{\gamma} \parallel F_{\gamma}$)
was nonempty, then $M_{\delta} \parallel T_{\delta}$ (respectively $M_{\delta} \parallel F_{\delta}$)
would have to be nonempty.
\qed
\end{proof}
We can now formally define the interpretation $M_P$ of a given program $P$: 
\[
       M_P(p) = \left\{
                    \begin{array}{ll}
                      M_{\delta}(p)& \mbox{if $order(M_{\delta}(p))<\delta$}\\
                      0 & \mbox{otherwise}
                    \end{array}
                  \right.
\]
As it will be shown shortly, $M_P$ is the least fixpoint of $T_P$, the minimum model 
of $P$ with respect to $\sqsubseteq_{\infty}$, and when it is restricted to three-valued 
logic it coincides with the well-founded model~\cite{vGRS91-620}.

\section{Properties of $M_P$}\label{properties}
In this section we demonstrate that the interpretation $M_P$ is a model of $P$.
Moreover, we show that $M_P$ is in fact the {\em minimum} model of $P$
under $\sqsubseteq_{\infty}$.
\begin{theorem}
The interpretation $M_P$ of a program $P$ is a fixpoint of $T_P$.
\end{theorem}
\begin{proof}
By the definition of $M_P$ and from Lemma \ref{emptylemma}, we have that for all 
countable ordinals $\alpha$ it is $M_P =_{\alpha} M_{\alpha}$. Then, for all $\alpha$, 
$T_P(M_P) =_{\alpha} T_P(M_{\alpha}) =_{\alpha} M_{\alpha} =_{\alpha} M_{P}$.
Therefore, $M_P$ is a fixpoint of $T_P$.
\qed
\end{proof}
\begin{theorem}\label{isamodel}
The interpretation $M_P$ of a program $P$ is a model of $P$.
\end{theorem}
\begin{proof}
Let $p \leftarrow B$ be a clause in $P$. It suffices to show that 
$M_P(p) \geq M_P(B)$.  We have:
\[
\begin{array}{llll}
M_P(p) &  =    & T_P(M_P)(p) & \mbox{(because $M_P$ is a fixpoint of $T_P$)}\\
       &  =    & lub\{M_P(B_C) \mid (p \leftarrow B_C) \in P \} & \mbox{(Definition of $T_P$)}\\
       &  \geq & M_P(B)      & \mbox{(Property of $lub$)}
\end{array}
\]
Therefore, $M_P$ is a model of $P$.
\qed
\end{proof}
The following lemma will be used in the proof of the main theorem of this section:
\begin{lemma}\label{inftylemma}
Let $N$ be a model of a given program $P$. Then, $T_P(N) \sqsubseteq_{\infty} N$.
\end{lemma}
\begin{proof}
Since $N$ is a model of $P$, then for all $p$ in $P$ and for all clauses of the 
form $p \leftarrow B$ in $P$, it is $N(p) \geq N(B)$. But then: 
$$T_P(N)(p) = lub\{N(B) \mid (p \leftarrow B) \in P \} \leq N(p)$$
Therefore, we have that $T_P(N)(p) \leq N(p)$ for all $p$ in $P$.
Using Lemma \ref{sqlemma}, we get that $T_P(N) \sqsubseteq_{\infty} N$.
\end{proof}
\begin{theorem}\label{minimum}
The infinite-valued model $M_P$ is the least (with respect to $\sqsubseteq_{\infty}$) 
among all infinite-valued models of $P$.
\end{theorem}
\begin{proof} 
Let $N$ be another model of $P$. We demonstrate that $M_P \sqsubseteq_{\infty} N$. It suffices
to show that for all countable ordinals $\alpha$, if for all $\beta < \alpha$
it is $M_P =_{\beta} N$ then $M_P \sqsubseteq_{\alpha} N$. The proof is by 
transfinite induction on $\alpha$. We distinguish three cases:

\vspace{0.2cm}

{\em Case 1}: $\alpha = 0$. We need to show that $M_P \sqsubseteq_0 N$. Now, since $M_P =_0 M_0$, 
it suffices to show that $M_0 \sqsubseteq_0 N$. By an inner induction, we demonstrate that 
for all $n<\omega$, $T_P^n(\emptyset)\sqsubseteq_0 N$. The basis case is trivial. Assume that 
$T_P^n(\emptyset)\sqsubseteq_0 N$. Using the $0$-monotonicity of $T_P$, we get that
$T_P^{n+1}(\emptyset) \sqsubseteq_0 T_P(N)$. From Lemma \ref{inftylemma} it is
$T_P(N) \sqsubseteq_{\infty} N$ which easily implies that
$T_P(N) \sqsubseteq_{0} N$. By the transitivity of $\sqsubseteq_{0}$ we get that 
$T_P^{n+1}(\emptyset) \sqsubseteq_0  N$. Therefore, for all $n<\omega$,
$T_P^n(\emptyset)\sqsubseteq_0 N$. Using Lemma \ref{lublemma} we get that
$M_0 \sqsubseteq_0 N$.

\vspace{0.2cm}

{\em Case 2:} $\alpha$ is a limit ordinal. We need to show that $M_P \sqsubseteq_{\alpha} N$.
Since $M_P =_{\alpha} M_{\alpha}$, it suffices to show that
$T_{P,\alpha}^{\omega}(\bigsqcup_{\beta < \alpha} M_{\beta})\sqsubseteq_{\alpha} N$.
This can be demonstrated by proving that for all $n < \omega$,
$T_{P}^{n}(\bigsqcup_{\beta < \alpha} M_{\beta}) \sqsubseteq_{\alpha} N$.
We proceed by induction on $n$. For $n=0$ the result is immediate. Assume the 
above statement holds for $n$. We need to demonstrate the statement for $n+1$. 
Using the $\alpha$-monotonicity of $T_P$, we get that
$T_{P}^{n+1}(\bigsqcup_{\beta < \alpha} M_{\beta})  \sqsubseteq_{\alpha} T_P(N)$.
Now, it is easy to see that for all $\beta < \alpha$, $T_P(N) =_{\beta} N$
(this follows from the fact that for all $\beta < \alpha$, $M_{\alpha} =_{\beta} N$).
From Lemma \ref{inftylemma} we also have $T_P(N) \sqsubseteq_{\infty} N$.
But then it is $T_P(N) \sqsubseteq_{\alpha} N$. Using the transitivity of $\sqsubseteq_{\alpha}$, 
we get that $T_{P}^{n+1}(\bigsqcup_{\beta < \alpha} M_{\beta}) \sqsubseteq_{\alpha} N$.
Therefore, for all $n < \omega$, $T_{P}^{n}(\bigsqcup_{\beta < \alpha} M_{\beta}) \sqsubseteq_{\alpha} N$.
Using Lemma \ref{lublemma} we get that $M_{\alpha} \sqsubseteq_{\alpha} N$.

\vspace{0.2cm}

{\em Case 3:} $\alpha$ is a successor ordinal. The proof is very similar to that 
for Case 2.
\end{proof}
\begin{corollary}
The infinite-valued model $M_P$ is the least (with respect to $\sqsubseteq_{\infty}$)
among all the fixpoints of $T_P$.
\end{corollary}
\begin{proof}
It is straightforward to show that every fixpoint of $T_P$ is a model of $P$ (the proof is identical 
to the proof of Theorem \ref{isamodel}). The result follows immediately since $M_P$ is the least model of $P$.
\qed
\end{proof}

Finally, the following theorem provides the connection between the infinite-valued semantics
and the existing semantic approaches to negation:
\begin{theorem}
Let $N_P$ be the interpretation that results from $M_P$ by collapsing all true values
to {\em True} and all false values to {\em False}. Then, $N_P$ is the well-founded
model of $P$.
\end{theorem}
\begin{proof}(Outline)
We consider the definition of the well-founded model given by T. Przymusinski in~\cite{P89-11}.
This construction uses three-valued interpretations but proceeds (from an algorithmic point
of view) in a similar way as the construction of the infinite-valued model. More specifically, the 
approximations of the well-founded model are defined in~\cite{P89-11} as follows (for a detailed explanation
of the notation, see~\cite{P89-11}):
\[
\begin{array}{ccll}
   M_0 & = & \langle T_{\emptyset},F_{\emptyset} \rangle & \\
   M_{\alpha} & = & M_{\alpha-1} \cup \langle T_{M_{\alpha-1}},F_{M_{\alpha-1}} \rangle & 
                               \mbox{for successor ordinal $\alpha$}\\
   M_{\alpha} & = & (\bigcup_{\beta < \alpha} M_{\beta}) \cup \langle T_{\bigcup_{\beta < \alpha} M_{\beta}},
                            F_{\bigcup_{\beta < \alpha} M_{\beta}} \rangle & 
                               \mbox{for limit ordinal $\alpha$}
\end{array}
\]
Notice that we have slightly altered the definition of~\cite{P89-11} for the case
of limit ordinals; the new definition leads to exactly the same model (obtained in 
a smaller number of steps). One can now show by a transfinite induction on $\alpha$ 
that the above construction introduces at each step exactly the same true and false
atoms as the infinite-valued approach.
\qed
\end{proof}

\section{A Model Intersection Theorem}\label{intersection}

In this section we demonstrate an alternative characterization of the minimum model $M_P$
of a program $P$. Actually, the proposed characterization generalizes the well-known 
{\em model intersection theorem} \cite{vanemden76,lloyd} that applies to classical logic programs 
(without negation).

The basic idea behind the model intersection theorem can be described as follows. 
Let $P$ be a given program and let ${\cal M}$ be the set of all its infinite-valued models.
We now consider all those models in ${\cal M}$ whose part corresponding to $T_0$ values
is equal to the intersection of all such parts for all models in ${\cal M}$, and whose part
corresponding to $F_0$ values is equal to the union of all such parts for all models in ${\cal M}$.
In other words, we consider all those models from ${\cal M}$ that have the fewest possible $T_0$ values and the
most $F_0$ values. This gives us a new set $S_0$ of models of $P$ (which as we demonstrate is non-empty).
We repeat the above procedure starting from $S_0$ and now considering values of order 1.
This gives us a new (non-empty) set $S_1$ of models of $P$, and so on. Finally, we demonstrate that
the limit of this procedure is a set that contains a unique model, namely the minimum model
$M_P$ of $P$. The above (intuitive) presentation can now be formalized as follows:
\begin{definition}
Let $S$ be a set of infinite-valued interpretations of a given program and $\alpha$ 
a countable ordinal. Then, we define 
$\bigwedge^{\alpha} S = \{(p,T_{\alpha}) \mid \forall M \in S, M(p) = T_{\alpha}\}$
and $\bigvee^{\alpha} S = \{(p,F_{\alpha}) \mid \exists M \in S, M(p) = F_{\alpha}\}$.
Moreover, we define $\bigodot^{\alpha} S = (\bigwedge^{\alpha} S) \bigcup (\bigvee^{\alpha} S)$.
\end{definition}
Let $P$ be a program and let ${\cal M}$ be the set of models of $P$. We can 
now define the following sequence of sets of models of $P$:
\[
\begin{array}{cclr}
   S_0        & = & \{M \in {\cal M} \mid M \sharp 0 = \bigodot^{0} {\cal M}\} & \\
   S_{\alpha} & = & \{M \in S_{\alpha - 1} \mid M \sharp \alpha = \bigodot^{\alpha}S_{\alpha - 1}\} 
                  &  \mbox{for successor ordinal $\alpha$}\\
   S_{\alpha} & = & \{M \in \bigcap_{\beta < \alpha} S_{\beta} \mid M \sharp \alpha = \bigodot^{\alpha} \bigcap_{\beta < \alpha} S_{\beta}\}
                  &  \mbox{for limit ordinal $\alpha$} 
\end{array}
\]
\begin{example}
Consider again the program of Example \ref{mainexample}:
\[
\begin{array}{lll}
 {\tt p} & \leftarrow & \mysim {\tt q}\\
 {\tt q} & \leftarrow & \mysim {\tt r}\\
 {\tt s} & \leftarrow & {\tt p}\\
 {\tt s} & \leftarrow & \mysim {\tt s} \\
 {\tt r} & \leftarrow & \mbox{\tt false}
\end{array}
\]
We first construct the set $S_0$. We start by observing that one of the models of the program
is the interpretation $\{({\tt r},F_0),({\tt q},T_1),({\tt p},F_2),({\tt s},0)\}$. Since this model does not contain any
$T_0$ value, we conclude that for all $M\in S_0$, $M\parallel T_0 = \emptyset$. Moreover, since the
above model contains $({\tt r},F_0)$, we conclude that for all $M \in S_0$, $({\tt r},F_0) \in M$. But this
implies that $({\tt q},T_1) \in M$, for all $M \in S_0$ (due to the second rule of the program and the 
fact that $M\parallel T_0 = \emptyset$). Using these restrictions, one can easily obtain restrictions
for the values of {\tt p} and {\tt s}. Therefore, the set $S_0$ consists of the following models:
\[
\begin{array}{lll}
S_0 & = & \{\{({\tt r},F_0),({\tt q},T_1),({\tt p},v_{\tt p}),({\tt s},v_{\tt s})\} \,\mid\, 
F_2 \leq v_{\tt p} \leq T_1,\, 0 \leq v_{\tt s} \leq T_1,\,u_{\tt s} \geq v_{\tt p}\}\\
\end{array}
\]
Now, observe that the model $\{({\tt r},F_0),({\tt q},T_1),({\tt p},F_2),({\tt s},0)\}$ belongs to $S_0$.
Since this model contains only one $T_1$ value, we conclude that for all 
$M\in S_1$, $M\parallel T_1 = \{{\tt q}\}$. Then, the set $S_1$ is the following:
\[
\begin{array}{lll}
S_1 & = & \{\{({\tt r},F_0),({\tt q},T_1),({\tt p},v_{\tt p}),({\tt s},v_{\tt s})\} \,\mid\, 
F_2 \leq v_{\tt p} \leq T_2, \, 0 \leq v_{\tt s} \leq T_2,\,u_{\tt s} \geq v_{\tt p}\}\\
\end{array}
\]
Using similar arguments as above we get that the set $S_2$ is the following:
\[
\begin{array}{lll}
S_2 & = & \{\{({\tt r},F_0),({\tt q},T_1),({\tt p},F_2),({\tt s},v_{\tt s})\} \,\mid\, 
0 \leq v_{\tt s} \leq T_3\}\\
\end{array}
\]
In general, given a countable ordinal $\alpha$, we have: 
\[
\begin{array}{lll}
S_{\alpha} & = & \{\{({\tt r},F_0),({\tt q},T_1),({\tt p},F_2),({\tt s},v_{\tt s})\} \,\mid\, 
0 \leq v_{\tt s} \leq T_{\alpha + 1}\}\\
\end{array}
\]
Observe that the model $\{({\tt r},F_0),({\tt q},T_1),({\tt p},F_2),({\tt s},0)\}$ is the only
model of the program that belongs to all $S_{\alpha}$. 
\end{example}

Consider now a program $P$ and let $S_0,S_1,\ldots,S_{\alpha},\ldots$ be the sequence of sets of
models of $P$ (as previously defined). We can now establish two lemmas that lead to the main theorem of
this section:
\begin{lemma}\label{nonempty}
For all countable ordinals $\alpha$, $S_{\alpha}$ is non-empty.
\end{lemma}
\begin{proof}
The proof is by transfinite induction on $\alpha$. We distinguish three cases:

\vspace{0.2cm}

\noindent
{\em Case 1:} $\alpha = 0$. Let $N^*$ be the following interpretation:
\[
       N^{*}(p) = \left\{
                    \begin{array}{ll}
                       T_{0}, &  \mbox{if $\forall M \in {\cal M}$ $(M(p) = T_{0})$}\\
                       F_{0}, &  \mbox{if $\exists M \in {\cal M}$ $(M(p) = F_{0})$}\\
                       T_{1}, & \mbox{otherwise}
                    \end{array}
                  \right.
\]
It is easy to show (by a case analysis on the value of $N^*(p)$) that $N^*$ is a model of program $P$
and therefore (due to the way it has been constructed) that $N^* \in S_0$.

\vspace{0.2cm}

\noindent
{\em Case 2:} $\alpha$ is a successor ordinal. Let $N \in S_{\alpha -1}$ be a model of $P$.
We construct an interpretation $N^*$ as follows:
\[
       N^{*}(p) = \left\{
                    \begin{array}{ll}
                       N(p),  & \mbox{if $order(N(p)) < \alpha$}\\
                       T_{\alpha}, &  \mbox{if $\forall M \in S_{\alpha -1}$ $(M(p) = T_{\alpha})$}\\
                       F_{\alpha}, &  \mbox{if $\exists M \in S_{\alpha -1}$ $(M(p) = F_{\alpha})$}\\
                       T_{\alpha +1}, & \mbox{otherwise}
                    \end{array}
                  \right.
\]
We demonstrate that $N^*$ is a model of $P$. Assume it is not. Then, there exists a clause
$p \leftarrow B$ in $P$ such that $N^*(p) < N^*(B)$. We perform a case analysis on the value 
of $N^*(p)$:
\begin{itemize}
\item $N^*(p) = F_{\beta}$, where $\beta \leq \alpha$. Then, there exists $M \in S_{\alpha -1}$ such that
      $M(p) = F_{\beta}$. Since $M$ is a model of $P$, for all clauses $p \leftarrow B_C$ in $P$, it is
      $M(B_C) \leq F_{\beta}$. Consequently, for every such clause, there exists a literal
      $l_C$ in $B_C$ such that $M(l_C) \leq F_{\beta}$. But then, it is also $N^*(l_C) \leq F_{\beta}$
      (by the definition of $N^*$ and since all models in $S_{\alpha - 1}$ agree on the values of order 
      less than $\alpha$). This implies that $N^*(B_C) \leq F_{\beta}$. Therefore, for all clauses of the form
      $p \leftarrow B_C$, it is $N^*(p) \geq N^*(B_C)$ (contradiction).

\item $N^*(p) = T_{\beta}$, $\beta \leq \alpha$. Since we have assumed that $N^*(p)<N^*(B)$, 
      it is $N^*(B) > T_{\beta}$. This implies that for every literal $l$ in $B$, it is 
      $N^*(l) > T_{\beta}$. But then, given any $M \in S_{\alpha -1}$, 
      it is also $M(l) > T_{\beta}$ (since all models in $S_{\alpha - 1}$ agree on
      the values of order less than $\alpha$). Therefore, $M(B) > T_{\beta}$. 
      But then, since $M(p) = T_{\beta}$, $M$ is not a model of $P$ (contradiction).

\item $N^*(p) = T_{\alpha + 1}$. Since we have assumed that $N^*(p)<N^*(B)$, it is 
      $N^*(B) \geq T_{\alpha}$. But then, for every $l \in B$, it is $N^*(l) \geq T_{\alpha}$. 
      Take now a model $M \in S_{\alpha - 1}$ such that $M(p)< T_{\alpha}$ 
      (such a model must exist because otherwise it would be $N^*(p) \geq T_{\alpha}$). 
      Now, it is easy to see that for every literal $l$ in $B$, 
      since it is $N^*(l) \geq T_{\alpha}$, it is $M(l) = N^*(l)$. This implies that 
      $M(B) \geq T_{\alpha}$. But since $M(p)< T_{\alpha}$, $M$ is not a 
      model of $P$ (contradiction).
\end{itemize}
Therefore, $N^*$ is a model of $P$. Moreover, due to the way it has been constructed,
$N^* \in S_{\alpha}$.

\vspace{0.2cm}

\noindent
{\em Case 3:} $\alpha$ is a limit ordinal. Let $N_0 \in S_{0},N_1 \in S_{1},\ldots,
N_{\beta} \in S_{\beta},\ldots,$ $\beta < \alpha$, be models of $P$. We construct an 
interpretation $N$ as follows:
\[
       N(p) = \left\{
                    \begin{array}{ll}
                      (\bigcup_{\beta < \alpha}(N_{\beta}\sharp \beta))(p)& \mbox{if this is defined}\\
                      T_{\alpha}& \mbox{otherwise}
                    \end{array}
                  \right.
\]
It is easy to see that $N$ is a model of $P$ and that $N\in \bigcap_{\beta < \alpha}S_{\beta}$.
This implies that the set $\bigcap_{\beta < \alpha}S_{\beta}$ is non-empty (which is needed in the 
definition that will follow). Now we can define an interpretation $N^{*}$ as follows:
\[
       N^{*}(p) = \left\{
                    \begin{array}{ll}
                       N(p),     & \mbox{if $order(N(p)) < \alpha$}\\
                       T_{\alpha}, &  \mbox{if $\forall M \in \bigcap_{\beta < \alpha}S_{\beta}$ $(M(p) = T_{\alpha})$}\\
                       F_{\alpha}, &  \mbox{if $\exists M \in \bigcap_{\beta < \alpha}S_{\beta}$ $(M(p) = F_{\alpha})$}\\
                       T_{\alpha +1}, & \mbox{otherwise}
                    \end{array}
                  \right.
\]
Then, using a proof very similar to the one given for Case 2 above, we can demonstrate
that $N^{*}$ is a model of $P$. Due to the way that it has been constructed, it is
obviously $N^{*} \in S_{\alpha}$. \qed
\end{proof}
\begin{lemma}\label{delta}
There exists a countable ordinal $\delta$ such that if $M \in S_{\delta}$ then:
\begin{enumerate}
\item   $M \sharp \delta = \emptyset$, and 
\item   for all $\gamma < \delta$, $M \sharp \gamma \neq \emptyset$.
\end{enumerate}
\end{lemma}
\begin{proof}
Since $B_P$ is countable, there can not be uncountably many $S_{\alpha}$ 
such that if $M \in S_{\alpha}$, $M \sharp \alpha \neq \emptyset$. Therefore,
we can take $\delta$ to be the smallest ordinal that satisfies the first 
condition of the lemma.
\end{proof}
We can now demonstrate the main theorem of this section which actually states
that there exists a unique model of $P$ that belongs to all $S_{\alpha}$:
\begin{theorem}
$\bigcap_{\alpha} S_{\alpha}$ is a singleton.
\end{theorem}
\begin{proof}
We first demonstrate that $\bigcap_{\alpha} S_{\alpha}$ can not contain more than one 
models. Assume that it contains two or more models, and take any two of them, say $N$ 
and $M$. Then, there must exist a countable ordinal, say $\gamma$, such that 
$N \sharp \gamma \neq M \sharp \gamma$. But then, $N$ and $M$ can not both belong to 
$S_{\gamma}$, and consequently they can not both belong to $\bigcap_{\alpha} S_{\alpha}$ 
(contradiction).

It remains to show that $\bigcap_{\alpha} S_{\alpha}$ is non-empty. By Lemma \ref{delta},
there exists $\delta$ such that if $M \in S_{\delta}$ then $M \sharp \delta = \emptyset$
(and for all $\gamma < \delta$, $M\sharp \gamma \neq \emptyset$).
Let $N \in S_{\delta}$ be a model (such a model exists because of Lemma \ref{nonempty}).
We can now create $N^{*}$ which is identical to $N$ but in which all atoms whose value 
under $N$ has order greater than $\delta$ are set to the value $0$. We demonstrate 
that $N^*$ is a model of the program. Assume it is not. Consider 
then a clause $p \leftarrow B$ such that $N^*(p) < N^*(B)$. There are three cases:
\begin{itemize}
\item $N^*(p) = F_{\beta}$, $\beta < \delta$. Then, $N(p) = F_{\beta}$ and since $N$
      is a model of $P$, we have $N(B) \leq F_{\beta}$. But this easily implies that
      $N^*(B) \leq F_{\beta}$, and therefore $N^*(p) \geq N^*(B)$ (contradiction).

\item $N^*(p) = T_{\beta}$, $\beta < \delta$. Then, $N(p) = T_{\beta}$ and since $N$
      is a model of $P$, we have $N(B) \leq T_{\beta}$. But this easily implies that
      $N^*(B) \leq T_{\beta}$, and therefore $N^*(p) \geq N^*(B)$ (contradiction).

\item $N^*(p) = 0$. Now, if $N(p) \leq 0$ then (since $N$ is a model) it is also $N(B) \leq 0$. 
      This easily implies that $N^{*}(B)\leq 0$. Therefore, $N^*(p) \geq N^*(B)$ (contradiction). 
      If on the other hand $N(p)>0$ then $N(p) < T_{\delta}$ (because $N^*(p) = 0$). Now, since 
      $N$ is a model, we have $N(B) < T_{\delta}$. But this easily implies that $N^*(B) \leq 0$ 
      and therefore $N^*(p) \geq N^*(B)$ (contradiction).
\end{itemize}
It is straightforward to see that (due to the way that it has been constructed)
$N^* \in S_{\alpha}$ for all countable ordinals $\alpha$. Therefore, 
$N^* \in \bigcap_{\alpha} S_{\alpha}$. \qed
\end{proof}

Finally, we need to establish that the model $M_P$ of $P$ produced through the $T_P$ operator
coincides with the model produced by the above theorem:
\begin{theorem}
$\bigcap_{\alpha} S_{\alpha} = \{M_P\}$
\end{theorem}
\begin{proof}
Let $N^*$ be the unique element of $\bigcap_{\alpha} S_{\alpha}$. Intuitively, due to the way
that it has been constructed, $N^*$ is ``as compact as possible'' at each level of truth values.
More formally, for every model $M$ of $P$ and for all countable ordinals $\alpha$,
if for all $\beta < \alpha$ it is $N^* =_{\beta} M$, then $N^* \sqsubseteq_{\alpha} M$ (the proof is
immediate due to the way that the sets $S_{\alpha}$ are constructed). Then, this implies
that $N^* \sqsubseteq_{\infty} M$. Take now $M$ to be equal to $M_P$. Then, $N^* \sqsubseteq_{\infty} M_P$
and also (from Theorem \ref{minimum}) $M_P \sqsubseteq_{\infty} N^{*}$. But since $\sqsubseteq_{\infty}$
is a partial order, we conclude that $N^* = M_{P}$. 
\qed
\end{proof}

\section{Discussion}\label{discussion}

In this section we argue (at an informal level) that the proposed
approach to the semantics of negation is closely related to the idea
of {\em infinitesimals} used in Nonstandard Analysis. Actually,
our truth domain can be understood as the result of extending the
classical truth domain by adding a neutral zero and a whole series of
{\em infinitesimal truth values} arbitrarily close to, but not equal 
to, the zero value.

Infinitesimals can be understood as values that are smaller
than any ``normal'' real number but still nonzero. In general,
each infinitesimal of order $n+1$ is considered to be infinitely smaller
than any infinitesimal of order $n$. It should be clear now how we can 
place our nonstandard logic in this context. We consider negation-as-failure as
ordinary negation followed by ``multiplication'' by an infinitesimal $\epsilon$.
$T_1$ and $F_1$ can be understood as the first order infinitesimals
$\epsilon T$ and $\epsilon F$, $T_2$ and $F_2$ as the second order
infinitesimals $\epsilon^2 T$ and $\epsilon^2 F$, and so on.

Our approach differs from the ``classical'' infinitesimals in that we
include infinitesimals of transfinite orders. Even in this respect,
however, we are not pioneers. John Conway, in his famous book
{\em On Numbers and Games}, constructs a field {\bf No} extending the 
reals that has infinitesimals of order $\alpha$ for {\em every} ordinal $\alpha$ - not 
just, as our truth domain, for every countable ordinal. Lakoff and Nunez give a 
similar (less formal) construction of what they call the {\em granular 
numbers}~\cite{lakoff}.  It seems, however, that we are the first to propose 
infinitesimal truth values.

But why are the truth values we introduced really infinitesimals?
Obviously $\epsilon T$ is smaller than $T$, $\epsilon^2 T$ is smaller 
than $\epsilon T$, and so on. But why are they infinitesimals - on what 
grounds can we claim that $\epsilon T$, for example, is {\em infinitely} smaller than $T$.
In the context of the real numbers, this question has a simple answer:
$\epsilon$ is infinitely smaller than $1$  because $n*\epsilon$ is 
smaller than $1$ for {\em any} integer $n$. Unfortunately, this formulation of the 
notion of ``infinitely smaller'' has no obvious analogue in logic because
there is no notion of multiplying a truth value by an integer.

There is, however, one important analogy with the classical theory of 
infinitesimals that emerges when we study the nonstandard ordering between models 
introduced.  Consider the problem of comparing two hyperreals each of which is
the sum of infinitesimals of different orders, ie. the problem of 
determining whether or not $A<B$, where $A = a_0+a_1*\epsilon+a_2*\epsilon^2+a_3*\epsilon^3+\cdots$
and $B = b_0+b_1*\epsilon+b_2*\epsilon^2+b_3*\epsilon^3+\cdots$
(with the $a_i$ and $b_i$ standard reals). We first compare $a_0$ and 
$b_0$.  If $a_0<b_0$ then we immediately conclude that $A<B$ without examining 
any other coefficients. Similarly, if $a_0>b_0$ then $A>B$. It is only in the case
that $a_0=b_0$ that the values $a_1$ and $b_1$ play a role. If they are 
unequal, $A$ and $B$ are ordered as $a_1$ and $b_1$. Only if $a_1$ and $b_1$ are 
{\em also} equal do we examine $a_2$ and $b_2$, and so on.

To see the analogy, let $I$ and $J$ be two of our nonstandard models 
and consider the problem of determining whether or not $I \sqsubseteq_{\infty} J$.  It 
is not hard to see that the formal definition of $I \sqsubseteq_{\infty}$ (given in Section 4)
can also be characterized as follows. First, let $I_0$ be the finite partial 
model which consists of the {\em standard} part of $I$ - the subset $I \parallel T_0  \cup  I \parallel F_0$ of $I$ 
obtained by restricting $I$ to those variables to which $I$ assigns standard truth values. Next, $I_1$ 
is the result of restricting $I$ to variables assigned order 1 infinitesimal values 
($T_1$ and $F_1$), and then replacing $T_1$ and $F_1$ by $T_0$ and $F_0$ (so that $I_1$ is 
also a standard interpretation). The higher ``coefficients''
$I_2, I_3,\ldots$ are defined in the same way. Then (stretching 
notation) $I = I_0 + I_1*\epsilon + I_2 * \epsilon^2 +\cdots$
and likewise $J = J_0 + J_1*\epsilon + J_2* \epsilon^2 +\cdots$.
Then to compare $I$ and $J$ we first compare the standard 
interpretations $I_0$ and $J_0$ using the standard relation. 
If $I_0 \sqsubseteq_0 J_0$, then $I \sqsubseteq_{\infty} J$. But
if $I_0 = J_0$, then we must compare $I_1$ and $J_1$, and if they are 
also equal, $I_2$ and $J_2$, and so on.  The analogy is actually very close, and 
reflects the fact that higher order truth values are negligible (equivalent to 0) 
compared to lower order truth values.

It seems that the concept of an infinitesimal truth value is closely 
related to the idea of prioritizing assertions. In constructing our minimal 
model the first priority is given to determining the values of the variables 
which receive standard truth values. This is the first approximation to the final 
model, and it involves essentially ignoring the contribution
of negated variables because a rule with negated variables in its body 
can never force the variable in the head of the clause to become $T_0$. In fact the whole 
construction proceeds according to a hierarchy of priorities corresponding to 
degrees of infinitesimals. This suggests that infinitesimal truth could be used in other 
contexts which seem to require prioritizing assertions, such as for example in default logic. 

\vspace{0.5cm}

\noindent
{\bf Acknowledgments:} We wish to thank Maarten van Emden, Bruce Kapron, Christos Nomikos
and John Schlipf for their comments on earlier versions of this paper. This work has been 
partially supported by the University of Athens under the project ``Extensions of the Logic 
Programming Paradigm'' (grant no. 70/4/5827).

\vspace{-0.2cm}

\bibliographystyle{alpha}
\bibliography{article,phdthesis,temporal,datalog,book,incollection,inproceedings}

\end{document}

\hyphenation{Dat-alog Data-log}
\newcommand{\skp}{\vskip 0.2cm }

\newcommand{\QED}{$\Box$}
\newcommand{\spa}{\,\,}
\newcommand{\pb}{\hspace*{3cm}}

\end{document}